\documentclass[runningheads]{llncs}
\usepackage[T1]{fontenc}
\usepackage{hyperref}
\usepackage{graphicx}
\usepackage{booktabs}
\usepackage{multirow}
\usepackage[skip=5pt]{caption}
\usepackage{color}

\usepackage{float}          
\usepackage{makecell}       
\usepackage{booktabs}       
\usepackage{multirow}       
\usepackage{array}          
\usepackage{graphicx}       
\usepackage{amsmath} 
\usepackage[style=ieee, maxnames=3, minnames=1]{biblatex}
\newcolumntype{T}[1]{>{\centering\arraybackslash}p{#1}}
\usepackage{multicol}
\newcolumntype{L}[1]{>{\raggedright\arraybackslash}m{#1}}

\begin{document}
\title{Segmenting infant brains across magnetic fields:\\Domain randomization and annotation curation in ultra-low field MRI}
\titlerunning{Segmenting infant brains across magnetic fields}




%
%

\author{Vladyslav Zalevskyi\inst{1,2} \and
Dondu-Busra Bulut\inst{1,2} \and\\
Thomas Sanchez\inst{1,2,*} \and
Meritxell Bach Cuadra\inst{1,2,*}\\[.3cm]
\email{vladyslav.zalevskyi@unil.ch}}
\authorrunning{V. Zalevskyi et al.}
%
\institute{Department of Radiology, Lausanne University Hospital and University of Lausanne (UNIL), Lausanne, Switzerland  \and
CIBM Center for Biomedical Imaging, Lausanne, Switzerland\\\textsuperscript{*}Shared senior authorship }

\maketitle              
\begin{abstract}
Early identification of neurodevelopmental disorders relies on accurate segmentation of brain structures in infancy, a task complicated by rapid brain growth, poor tissue contrast, and motion artifacts in pediatric MRI. These challenges are further exacerbated in ultra-low-field (ULF, 0.064~T) MRI, which, despite its lower image quality, offers an affordable, portable, and sedation-free alternative for use in low-resource settings. In this work, we propose a domain randomization (DR) framework to bridge the domain gap between high-field (HF) and ULF MRI in the context of the hippocampi and basal ganglia segmentation in the LISA challenge. We show that pre-training on whole-brain HF segmentations using DR significantly improves generalization to ULF data, and that careful curation of training labels, by removing misregistered HF-to-ULF annotations from training, further boosts performance. By fusing the predictions of several models through majority voting, we are able to achieve competitive performance. Our results demonstrate that combining robust augmentation with annotation quality control can enable accurate segmentation in ULF data. Our code is available at \url{https://github.com/Medical-Image-Analysis-Laboratory/lisasegm}

\end{abstract}

\section{Introduction}

\looseness=-1
Accurate segmentation of brain structures in early childhood is essential for studying typical neurodevelopment and identifying early signs of neurological disorders. Deep gray matter regions such as the basal ganglia and hippocampi are particularly important due to their roles in motor control, cognition, and memory—functions commonly impacted in conditions like ADHD and autism spectrum disorders~\cite{barnea2014preliminary,hoogman2017subcortical,xu2020abnormal}. However, existing deep learning segmentation models, largely trained on adult high-field (HF) MRI, often perform poorly on pediatric low-field (LF, 0.1-1\,T) and ultra-low-field (ULF, $<$0.1T) scans. This generalization gap stems from anatomical differences between adults and infants, field-strength–dependent contrast changes, and frequent motion artifacts in pediatric populations~\cite{lisa25zenodo}.  These challenges are amplified in low-resource settings, where access to HF scanners (1.5\,T/3\,T) is limited~\cite{jalloul2023mri,murali2024bringing}. ULF MRI systems, such as the 0.064\,T Hyperfine scanner, offer a portable and cost-effective alternative for pediatric imaging~\cite{Iglesias2023qMRIULF,zhao2024whole,Johnson2025,gopinath2025low}. Yet, their lower signal-to-noise ratio and reduced spatial resolution demand methods that are explicitly adapted to this imaging regime.

Domain randomization (DR) has emerged as a promising strategy to improve generalization across MRI modalities, contrasts, and populations. Methods such as \texttt{SynthSeg}~\cite{billot2023synthseg,billot2023robust} augment label-derived images with randomized intensity and spatial transformations, enabling robust performance even on highly heterogeneous or low-quality data~\cite{vavsa2024ultra}. In particular, DR has recently been applied to fetal and neonatal imaging~\cite{valabregue2024comprehensive,zalevskyi2024maximizing}, but to our knowledge, no prior work has addressed the joint challenge of pediatric anatomy and ULF domain shift.

In this paper, we investigate DR-based pre-training for segmenting deep gray matter structures in infant ULF MRI, as part of the LISA challenge. We evaluate how DR can support cross-domain generalization from HF data and examine the impact of annotation quality on fine-tuning performance. Specifically, we show that: \textbf{(1)}~Domain randomization enables HF-to-ULF annotation transfer, \textbf{(2)}~Pre-training on whole-brain annotations boosts task-specific segmentation, \textbf{(3)}~Filtering out misaligned labels improves robustness, and \textbf{(4)}~Model ensembling further increases the performance of our models.

We believe that these observations will inform the design of segmentation pipelines that are both accurate and practical, particularly in low-resource clinical environments where ULF MRI offers a promising solution for early neurodevelopmental assessment.


\section{Methods}
\subsection{Datasets}
We used three datasets with neonatal and infant brain images and segmentations in this work. Two publicly available HF datasets were used for pre‑training (dHCP~\cite{Edwards2022dHCP} and BOBs~\cite{Feczko2024BOBs}), and the LISA challenge dataset~\cite{lisa25zenodo} is used for fine‑tuning. A summary is provided in Table~\ref{tab:datasets}.

\begin{table}[t]
    \centering
    \includegraphics[width=1\linewidth]{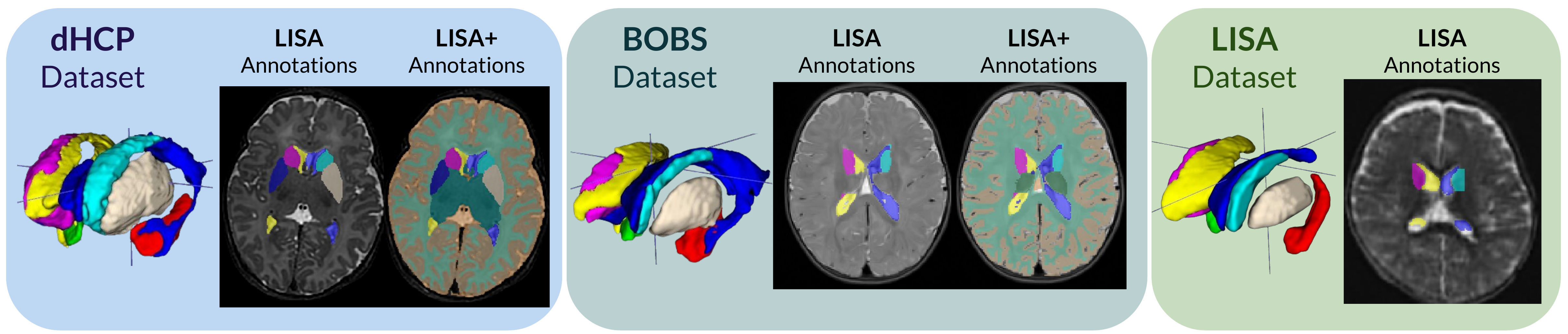}
    \captionof{figure}{Overview of the datasets and annotation schemes used in this study. Some variations are found across dataset annotation definition: e.g. in both dHCP and BOBs the lateral ventricle label extends further down toward the hippocampi).}
    \label{fig:dataoverview}

\captionof{table}{Summary of the datasets used for pre-training (dHCP and BOBs) and fine-tuning (LISA challenge dataset).}
\label{tab:datasets}
\renewcommand{\arraystretch}{1.3}
\resizebox{.9\linewidth}{!}{
\begin{tabular}{
m{1.8cm}  
L{2cm}   
L{2.5cm} 
L{2.2cm} 
L{2.3cm}   
m{1.8cm} 
}
\toprule
\textbf{Dataset} & \textbf{Scanner} & \textbf{Age range} & \textbf{Conditions} & \textbf{Size} & \textbf{Resolution} \\
\midrule
\textbf{BOBs}~\cite{Feczko2024BOBs} & 3T Siemens Prisma & 1--9~months & Healthy & 51 infants, 71 sessions & 0.8~mm$^3$ \\
\textbf{dHCP}~\cite{Edwards2022dHCP} & 3T Philips Achieva & 23--44 gestational weeks & Healthy & 783 infants, 887 sessions & 0.5~mm$^3$ \\
\midrule[2pt]
\textbf{LISA}~\cite{lisa25zenodo} & 0.064T Hyperfine SWOOP & 4.5~weeks--16~months & Healthy & 79 (training), 12 (validation) & 1~mm$^3$ \\
\bottomrule
\end{tabular}}
\vspace{-.3cm}
\end{table}

\noindent\textbf{Baby Open Benchmark Segmentations (BOBs)}
The BOBs dataset~\cite{feczko2024baby} was acquired on a high‑field 3T Siemens Prisma system at the University of Minnesota, United States, following a Baby Connectome Project (BCP) protocol~\cite{Howell2019}. This subset includes only healthy subjects for which extensive, manually curated segmentations are available. The segmentations are aligned to both T1‑weighted and T2‑weighted scans that were acquired and include labels for cerebral gray matter, white matter, and 23 subcortical structures. 

\noindent\textbf{Developing Human Connectome Project (dHCP)}
The dHCP dataset~\cite{edwards2022developing} was acquired at the Evelina Newborn Imaging Centre, King’s College London, using a 3T Philips Achieva scanner optimized for neonatal imaging. We used the T2-weighted images with the provided automated segmentations into 9 tissue classes and 87 regions~\cite{Makropoulos2018}. Note that while the data is rich, the dHCP dataset is centered around the perinatal period, compared to the LISA challenge data covering subjects up to 16 months.

\noindent\textbf{LISA dataset}
The Low‑field Pediatric Brain Magnetic Resonance Image Segmentation and Quality Assurance (LISA) dataset~\cite{lisa25zenodo} was collected from healthy neonates at three sites: the University of Cape Town (South Africa), Makerere University (Uganda), and Aga Khan University Hospital (Pakistan). Imaging was performed using portable ultra‑low‑field (ULF) 0.064~T Hyperfine SWOOP scanners. For this work, we tackled the second task of the LISA challenge, which consisted in segmenting super‑resolution‑reconstructed~\cite{deoni2022simultaneous} T2‑weighted images across 8 structures: left and right hippocampi, lateral ventricles, caudate nuclei, and lentiform nuclei. Two sets of ground truth labels were provided: initial labels done on the ULF images (\texttt{GT}$_\text{LF}$), and labels created on the high‑field (1.5T or 3T) image counterparts and subsequently propagated to the ULF images through nine‑point linear co‑registration (\texttt{GT}$_\text{HF}$). The second set of labels was the one used as ground truth in the challenge.

\subsection{Data harmonization and pre-processing}
We first resampled every dataset’s images to an isotropic voxel size of 1 $mm^3$, matching the resolution of the LISA dataset. 
Because the pre-training datasets (dHCP and BOBs) contained a much denser set of annotation labels than those defined in LISA, we pre-trained our models using two annotation variants, as illustrated in Figure~\ref{fig:dataoverview}:
\begin{enumerate}
    \item \textbf{LISA:} We selected from the pre-training datasets only the labels present in the LISA dataset. Both datasets contain the required classes for the LISA challenge, although with slightly different definitions.
    \item \textbf{LISA}$+$\textbf{:} We used the dense structural information from dHCP and BOBs to pre-train our models. As dHCP and BOBs contained many labels, we grouped their annotations in 6 groups covering the entire brain:  white matter (WM), cortical gray matter (GM), cerebrospinal fluid (CSF -- dHCP only), cerebellum, brainstem, and deep GM (excluding the basal ganglia and hippocampi). These labels were used in addition to the 8 LISA labels during training. 
\end{enumerate}

\vspace{-.3cm}
\subsection{Domain randomization}

In this work, we used domain randomization as the core of our training procedure. DR relies on a generating procedure producing randomized intensity images with paired annotations that are then used to train a segmentation model. 

\subsubsection{Data generation.}
For the generation procedure, we used the synthetic data generator from the \texttt{FetalSynthSeg} framework by Zalevskyi \textit{et al.}~\cite{zalevskyi2024maximizing} and adapted it to the ULF setting. This publicly available generator\footnote{\url{https://github.com/Medical-Image-Analysis-Laboratory/fetalsyngen}} was tailored for fetal populations using T2w images and was easy to adapt to the pediatric population of the LISA challenge. 

Most parameters of the generator were kept unchanged: rigid and non-rigid spatial deformations, intensity and contrast randomization, as well as intensity generation were kept identical to previous work on low‑field (0.55 T) fetal brain segmentation~\cite{zalevskyi2024maximizing} as we found these parameters to readily apply to the ULF domain. We changed the resolution resampling step to downsample the generated volume to an anisotropic volume with slice thickness between 1 and 5~mm, before interpolating it back to an isotropic resolution of 1~mm$^3$ (simulating reconstruction of ULF images). 
To account for ULF‑specific artifacts not simulated by \texttt{FetalSynthSeg}, we additionally applied a set of randomized augmentations that simulate common artifacts in ULF MRI: random k-space motion artifacts, k‑space ghosting, and spiking artifacts~\cite{perez2021torchio}. These artifacts improved the robustness of our model---even though it remains vulnerable to artifacts not simulated during training---while the domain randomized approach allowed it to perform well even on cases with very low tissue contrast.

\begin{figure}[t]
\vspace{-.3cm}
    \centering
    \includegraphics[width=\linewidth]{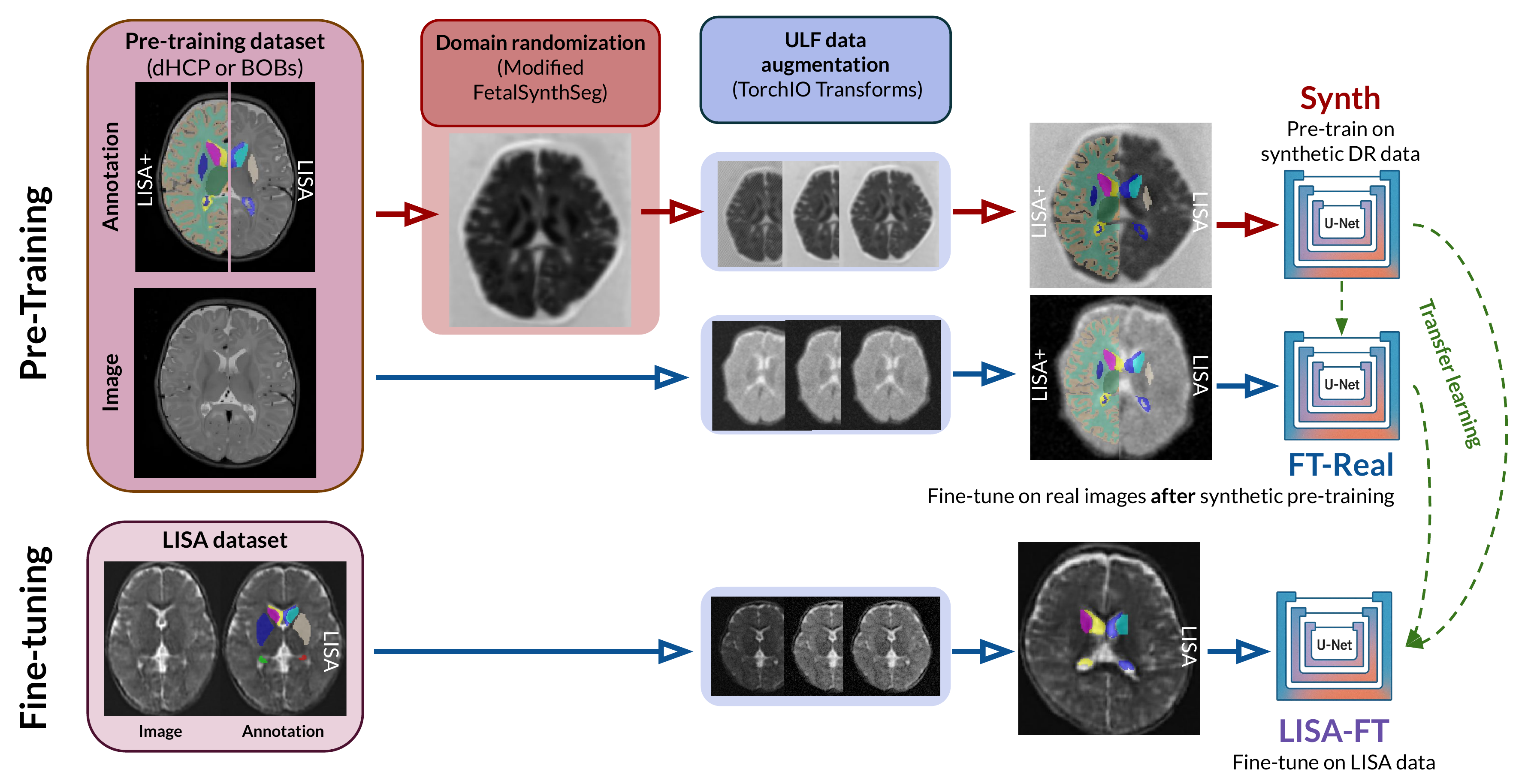}
    \caption{Overview of training modes used in this study. We start with a pre-training step yielding a \texttt{Synth} model that can then be fine-tuned on the real images from the pre-training data to yield \texttt{FT-Real}. These models are then fine-tuned on LISA data for evaluation. Red arrows show the use of synthetic data, blue arrows show real images, and green arrows show fine-tuning paths.}
    \label{fig:trainingflow}
    \vspace{-.3cm}
\end{figure}

\vspace{-.3cm}
\subsubsection{Model pre-training.}
\looseness=-1
Using the generated data and corresponding labels, we pre-trained two model variants on the dHCP and BOBs datasets: \textbf{(i) \texttt{Synth}:} a model trained only on synthetic data, and \textbf{(ii) \texttt{FT-Real}:} a model trained on synthetic data and fine-tuned on the real images from the pre-training dataset, as this was found to yield improved performance in previous work~\cite{zalevskyi2024maximizing}. We did not leverage a mix of real and synthetic data in training, as this approach generally worsens performance by going against the tested philosophy of domain randomization: maximizing training variability to be robust to changes at deployment time.  
The fine-tuning step also included the randomized resolution resampling and ULF augmentations mentioned above. The training pipeline is summarized in Figure~\ref{fig:trainingflow}. 

\vspace{-.3cm}
\subsubsection{Model fine-tuning.}
The pre-trained models were then fine-tuned on the LISA dataset, on all target labels at once. When pre-trained models were segmenting the entire brain, we fine-tuned the model by only optimizing the output channels relevant to our tasks ( hippocampi and basal ganglia labels). This step followed the same training parameters as above, with the only exception that we did not use randomized resolution resampling here. After fine-tuning, we also explored various model combinations using ensembling through majority voting~\cite{eisenmann2023winner}.

\vspace{-.3cm}
\subsubsection{Model architecture and optimization parameters}
All individual models we train in our experiments were based on a 3D U-Net implemented in MONAI~\cite{monai}. The network started with 32 feature channels and doubled the number of channels at each downsampling stage. Convolutional layers used $3 \times 3 \times 3$ kernels with LeakyReLU activations, and the final layer applied a softmax function. Skip connections were included between corresponding encoder and decoder levels. Training was performed with the Adam optimizer (learning rate $10^{-3}$) using a combined Dice–Cross-Entropy loss. To ensure stable convergence, we employed a \texttt{ReduceLROnPlateau} scheduler (factor $0.1$, patience $=10$) and early stopping (patience $=100$ iterations). During training, we reserved 10\% of the available training data as an internal validation set, which was used to monitor performance and trigger early stopping. All experiments were run on NVIDIA RTX~6000 GPUs using PyTorch Lightning with a batch size of 1. ULF augmentations were generated using TorchIO~\cite{perez2021torchio}. Upon acceptance of the paper, we will release the full training and image‑generation code as well as the best model weights for reproducibility.

\vspace{-.3cm}
\subsection{Experiment setting}
We conducted several rounds of experiments to evaluate our hypotheses. All models were assessed on the LISA challenge validation dataset, and we reported the official metrics published on the challenge website on the day of the paper submission. In all experiments, the models were optimized to segment the whole set of 8 labels provided in the LISA challenge, thus enabling a joint optimization for both sub-tasks 2a and 2b simultaneously. 

\vspace{-.3cm}
\subsubsection{Metrics.} 
\looseness=-1
We reported the metrics included in the evaluation of LISA challenge\cite{lisa25zenodo}, namely Dice Score (DSC; $\uparrow$), Hausdorff  distance (HD; $\downarrow$),  95\% HD (HD95; $\downarrow$), average symmetric surface distance (ASSD; $\downarrow$), and relative volume error (RVE; $\downarrow$). The metrics covered only the labels evaluated in the scope of the challenge (e.g., excluding lateral ventricles). In addition, we also computed the \textit{final challenge ranking metric}, which was computed as the mean of the normalized scores of all other metrics\footnote{Using 1-DSC to reverse the metric value and align its scale to others.}, each normalized between the worst and best values across all submissions. For simplicity, we refer to this aggregated normalized averaged metric as \textbf{\texttt{NormAvg}} throughout the results.

\subsubsection{Experiments.} Our experiments explored three questions.

\noindent\textbf{1. Can DR help leverage external HF MRI datasets of infants/neonates for ULF segmentation?} 
\looseness=-1
For each combination of pre-training dataset (BOBs or dHCP) and annotation scheme (LISA or LISA$_{+}$), we trained and evaluated the \texttt{Synth} and \texttt{FT-Real} variants of our models.
We then carried out a qualitative evaluation to assess the out-of-the-box performance of our models pre-trained on the high-field dHCP and BOBs datasets. During this step, we observed some misregistered labels in the LISA data. This prompted us to carry out a manual \textit{quality assessment}: we looked at the right ventricle and caudate nucleus across the data and rated as \texttt{bad} the samples where these structures were misregistered during label propagation from HF segmentations.

\noindent\textbf{2. What is the most efficient pre-training for each task?}
In this experiment, we fine-tuned the different pre-trained models of experiment 1 on the LISA dataset and evaluated them quantitatively. We also explored how label quality impacted the fine-tuning performance of these models, using either \texttt{all} data, only \texttt{good} data, or only misregistered data (denoted as \texttt{bad} for conciseness).

\noindent\textbf{3. How to get the most out of ensembling?}  
Finally, we explored the combination of various fine-tunings of our pre‑trained models with different annotation schemes, through a voxel‑wise majority voting.

\section{Results and discussion}

\subsection{Domain randomization bridges HF and ULF annotations}
\looseness=-1
Table~\ref{tab:resultspretrain} summarizes the results of pre‑training the models on dHCP and BOBs datasets and their direct evaluation on LISA (no training on LISA data).
\begin{table}[t]
  \centering
  \scriptsize
  \caption{Pre-training results on LISA validation dataset. Best value in each training group is in \textbf{bold} (\texttt{NormAvg}. only).}
  \label{tab:resultspretrain}
  \resizebox{.9\linewidth}{!}{
  \begin{tabular}{ccllcccccc}
    \toprule
    Task & Pretrain. & Annotation & Training & \texttt{NormAvg} $\downarrow$ & DSC $\uparrow$ & HD $\downarrow$ & HD95 $\downarrow$ & ASSD $\downarrow$ & RVE $\downarrow$ \\
    \midrule

    \multirow{8}{*}{\rotatebox{90}{\hspace{-.3cm}\textsc{2a--Hippocampus}}} 
      & \multirow{4}{*}{\rotatebox{90}{\textsc{BOBs}}} 
        & \multirow{2}{*}{LISA}    
          & \texttt{FT‑Real} & \textbf{1.385} & 0.56{\tiny ±0.16} & 12.46{\tiny ±7.92} & 6.89{\tiny ±1.34} & 1.52{\tiny ±0.88} & 0.28{\tiny ±0.18} \\
      &      &        & \texttt{Synth}   & 1.610           & 0.54{\tiny ±0.14} & 10.96{\tiny ±6.97} & 5.92{\tiny ±1.55} & 1.61{\tiny ±0.75} & 0.60{\tiny ±0.25} \\
    \cmidrule[0.1pt]{3-10}
      &      & \multirow{2}{*}{LISA$_{+}$}  
          & \texttt{FT‑Real} & \textbf{1.240} & 0.57{\tiny ±0.14} & 9.97{\tiny ±7.35}  & 4.89{\tiny ±1.30} & 1.40{\tiny ±0.76} & 0.45{\tiny ±0.22} \\
      &      &        & \texttt{Synth}   & 1.809           & 0.52{\tiny ±0.13} & 11.16{\tiny ±7.19} & 5.94{\tiny ±1.47} & 1.67{\tiny ±0.62} & 0.74{\tiny ±0.32} \\

    \cmidrule[0.1pt]{2-10}

      & \multirow{4}{*}{\rotatebox{90}{\textsc{dHCP}}} 
        & \multirow{2}{*}{LISA}    
          & \texttt{FT‑Real} & \textbf{3.532} & 0.32{\tiny ±0.08} & 18.34{\tiny ±6.98} & 12.49{\tiny ±1.69} & 3.25{\tiny ±0.55} & 0.91{\tiny ±0.42} \\
      &      &        & \texttt{Synth}   & 4.169           & 0.25{\tiny ±0.06} & 19.87{\tiny ±6.65} & 14.18{\tiny ±2.19} & 3.98{\tiny ±0.67} & 1.06{\tiny ±0.42} \\
    \cmidrule[0.1pt]{3-10}
      &      & \multirow{2}{*}{LISA$_{+}$} 
          & \texttt{FT‑Real} & 2.881           & 0.38{\tiny ±0.09} & 18.16{\tiny ±7.52} & 12.35{\tiny ±2.52} & 2.93{\tiny ±0.84} & 0.45{\tiny ±0.28} \\
      &      &        & \texttt{Synth}   & \textbf{2.841}           & 0.40{\tiny ±0.09} & 16.75{\tiny ±6.87} & 10.89{\tiny ±1.56} & 2.71{\tiny ±0.53} & 0.68{\tiny ±0.32} \\

    \midrule

    \multirow{8}{*}{\rotatebox{90}{\hspace{-.3cm}\textsc{2b--Basal ganglia}}}  
      & \multirow{4}{*}{\rotatebox{90}{\textsc{BOBs}}} 
        & \multirow{2}{*}{LISA}    
          & \texttt{FT‑Real} & \textbf{1.011} & 0.76{\tiny ±0.04} & 5.83{\tiny ±1.11}  & 3.52{\tiny ±1.08}  & 0.92{\tiny ±0.23} & 0.16{\tiny ±0.06} \\
      &      &        & \texttt{Synth}   & 1.088           & 0.74{\tiny ±0.05} & 6.08{\tiny ±1.06}  & 3.65{\tiny ±0.98}  & 0.99{\tiny ±0.23} & 0.14{\tiny ±0.10} \\
    \cmidrule[0.1pt]{3-10}
      &      & \multirow{2}{*}{LISA$_{+}$} 
          & \texttt{FT‑Real} & 1.469           & 0.71{\tiny ±0.05} & 6.68{\tiny ±1.56}  & 4.50{\tiny ±1.20}  & 1.13{\tiny ±0.26} & 0.22{\tiny ±0.06} \\
      &      &        & \texttt{Synth}   & \textbf{1.460}           & 0.73{\tiny ±0.05} & 7.02{\tiny ±1.02}  & 4.52{\tiny ±1.04}  & 1.11{\tiny ±0.21} & 0.24{\tiny ±0.10} \\

    \cmidrule[0.1pt]{2-10}

      & \multirow{4}{*}{\rotatebox{90}{\textsc{dHCP}}} 
        & \multirow{2}{*}{LISA}    
          & \texttt{FT‑Real} & \textbf{3.669} & 0.53{\tiny ±0.04} & 10.91{\tiny ±1.06} & 8.03{\tiny ±1.09}  & 2.20{\tiny ±0.28} & 0.68{\tiny ±0.20} \\
      &      &        & \texttt{Synth}   & 4.070           & 0.50{\tiny ±0.05} & 11.83{\tiny ±1.37} & 9.10{\tiny ±1.15}  & 2.45{\tiny ±0.27} & 0.72{\tiny ±0.20} \\
    \cmidrule[0.1pt]{3-10}
      &      & \multirow{2}{*}{LISA$_{+}$} 
          & \texttt{FT‑Real} & \textbf{3.915}           & 0.57{\tiny ±0.05} & 11.66{\tiny ±4.01} & 8.53{\tiny ±4.04}  & 3.01{\tiny ±4.00} & 0.66{\tiny ±0.16} \\
      &      &        & \texttt{Synth}   & 4.274           & 0.51{\tiny ±0.08} & 12.37{\tiny ±3.67} & 9.66{\tiny ±3.85}  & 3.23{\tiny ±3.36} & 0.63{\tiny ±0.20} \\

    \bottomrule
  \end{tabular}}
  \vspace{-.5cm}
\end{table}

\begin{figure}[t]
    \centering
    \includegraphics[width=1\linewidth]{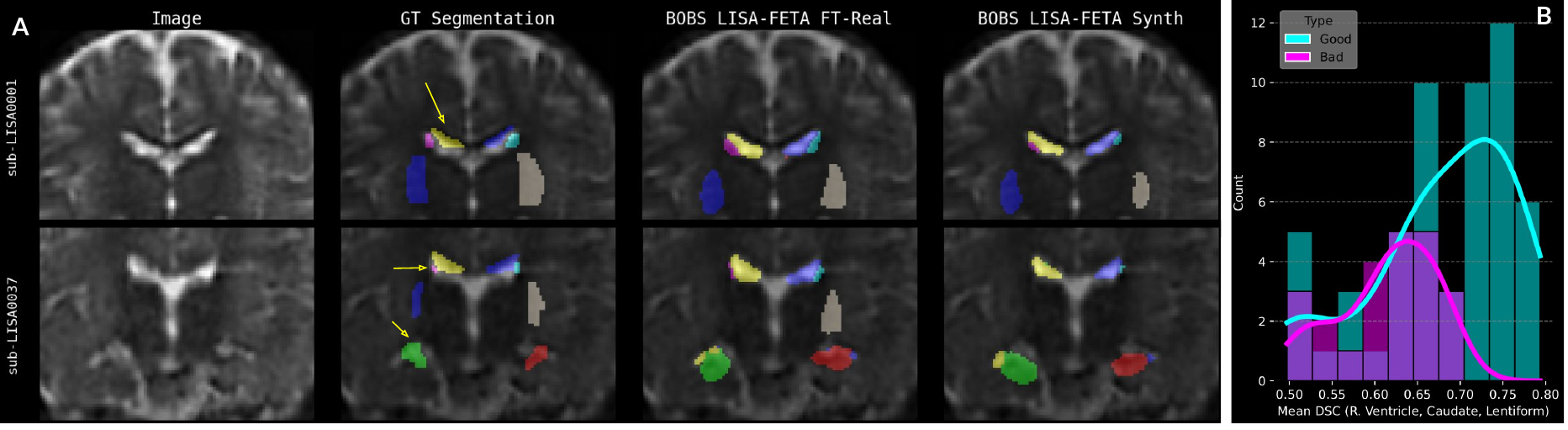}
    \caption{
\textbf{(A)} Examples of annotation drift caused by registration errors in the ground‑truth (GT) segmentations of two representative LISA cases, compared with the outputs from our pre‑trained models.
\textbf{(B)} Distribution of the mean DSC for several critically affected right‑sided labels across the training dataset, comparing the LISA GT annotations with the BOBs +LISA$_{+}$+\texttt{FT‑Real} model. Note the bimodal distribution, reflecting the difference in Dice scores between well‑registered and poorly registered images.  
The images shown in panel (A) have mean DSC values of 0.53 and 0.57, respectively.
}
    \label{fig:badannot}

\centering

\captionof{table}{Quantitative metrics by task, annotation, and training type (training on exclusively images that we deemed \texttt{good} or \texttt{all} training data). All models trained are on BOBs dataset.}
\label{tab:goodbadall}
\resizebox{.8\linewidth}{!}{

\begin{tabular}{c l l l c c c c c c}
\toprule
\rotatebox{0}{Task}
  & \rotatebox{0}{Annot.}
    & \rotatebox{0}{Training}
      & Type      & \texttt{NormAvg} $\downarrow$ & DSC $\uparrow$ & HD $\downarrow$ & HD95 $\downarrow$ & ASSD $\downarrow$ & RVE $\downarrow$      \\
\midrule
\multirow{6}{*}{\rotatebox{90}{\hspace{-0.3cm}\textsc{2a--Hippocampus}}}
  & \multirow{2}{*}{LISA}
    & \texttt{FT-Real} & All & 0.268
              & 0.67{\tiny ±0.18}
              & 6.41{\tiny ±8.21}
              & 2.45{\tiny ±1.24}
              & 0.86{\tiny ±0.81}
              & 0.17{\tiny ±0.08} \\
  &            & \texttt{FT-Real} & Good & \textbf{0.215}
              & 0.68{\tiny ±0.19}
              & 6.16{\tiny ±8.05}
              & 2.25{\tiny ±1.67}
              & 0.87{\tiny ±1.08}
              & 0.16{\tiny ±0.07} \\
\cmidrule(lr){2-10}
  & \multirow{2}{*}{LISA}
    & \texttt{Synth} & All & \textbf{0.183}
              & 0.70{\tiny ±0.19}
              & 6.06{\tiny ±7.74}
              & 2.23{\tiny ±1.56}
              & 0.81{\tiny ±0.95}
              & 0.18{\tiny ±0.07} \\
  &            & \texttt{Synth} & Good & 0.231
              & 0.69{\tiny ±0.18}
              & 6.27{\tiny ±8.06}
              & 2.31{\tiny ±1.64}
              & 0.85{\tiny ±1.05}
              & 0.19{\tiny ±0.07} \\
\cmidrule(lr){2-10}
  & \multirow{2}{*}{LISA$_{+}$}
    & \texttt{FT-Real} & All & 0.199
              & 0.68{\tiny ±0.18}
              & 5.83{\tiny ±7.88}
              & 2.15{\tiny ±1.28}
              & 0.80{\tiny ±0.77}
              & 0.17{\tiny ±0.07} \\
  &            & \texttt{FT-Real} & Good & \textbf{0.118}
              & 0.69{\tiny ±0.18}
              & 5.73{\tiny ±7.79}
              & 2.02{\tiny ±1.42}
              & 0.80{\tiny ±0.90}
              & 0.12{\tiny ±0.08} \\
\midrule
\multirow{6}{*}{\rotatebox{90}{\hspace{-0.3cm}\textsc{2b--BaGa}}}
  & \multirow{2}{*}{LISA}
    & \texttt{FT-Real} & All & 0.150
              & 0.85{\tiny ±0.05}
              & 3.17{\tiny ±0.86}
              & 1.76{\tiny ±0.93}
              & 0.49{\tiny ±0.21}
              & 0.07{\tiny ±0.02} \\
  &            & \texttt{FT-Real} & Good & \textbf{0.135}
              & 0.86{\tiny ±0.07}
              & 3.20{\tiny ±1.05}
              & 1.75{\tiny ±0.93}
              & 0.48{\tiny ±0.30}
              & 0.08{\tiny ±0.04} \\
\cmidrule(lr){2-10}
  & \multirow{2}{*}{LISA}
    & \texttt{Synth} & All & \textbf{0.129}
              & 0.86{\tiny ±0.04}
              & 3.04{\tiny ±0.69}
              & 1.64{\tiny ±0.69}
              & 0.48{\tiny ±0.21}
              & 0.09{\tiny ±0.03} \\
  &            & \texttt{Synth} & Good & 0.236
              & 0.84{\tiny ±0.07}
              & 3.17{\tiny ±0.96}
              & 1.89{\tiny ±1.00}
              & 0.53{\tiny ±0.29}
              & 0.09{\tiny ±0.04} \\
\cmidrule(lr){2-10}
  & \multirow{2}{*}{LISA$_{+}$}
    & \texttt{FT-Real} & All & \textbf{0.182}
              & 0.85{\tiny ±0.06}
              & 3.20{\tiny ±0.99}
              & 1.76{\tiny ±0.98}
              & 0.49{\tiny ±0.26}
              & 0.09{\tiny ±0.03} \\
  &            & \texttt{FT-Real} & Good & 0.196
              & 0.85{\tiny ±0.07}
              & 3.28{\tiny ±1.19}
              & 1.89{\tiny ±1.05}
              & 0.52{\tiny ±0.29}
              & 0.08{\tiny ±0.03} \\
\bottomrule
\end{tabular}}
\vspace{-.4cm}
\end{figure}

Across most experiments, \texttt{FT‑Real} models (trained on synthetic data and then fine‑tuned on real high‑field images) consistently outperformed those trained on synthetic data alone. Despite the HF–ULF domain shift, the models benefited from additional T2 information in the HF data. Having never seen the LISA data, these models were not competitive quantitatively, but BOBs-based models still yielded qualitatively good segmentations, especially in the ventricles and caudate nuclei, as illustrated in Figure~\ref{fig:badannot}. This figure also helped us realize that some of the ground truth annotations were \textit{misregistered} (see yellow arrows on the left side of the image). We also observed that the segmentations produced by our pre‑trained models were well aligned in these cases.

A manual inspection of the provided training dataset revealed that the HF annotations had some misalignment around the right hemisphere ventricle and caudate in at least 23 out of 79 training cases, similar to those errors depicted in the figure \ref{fig:badannot} A. On Figure~\ref{fig:badannot} B, the Dice score computed on these structures using our pre-trained models and the ground truth LISA labels did show a correlation between a lower Dice score and data that we rated as misaligned. A more detailed explanation of our manual label quality rating is provided in Appendix~\ref{app:qc}.

\subsection{The effect of annotation quality on model training}
\looseness=-1
To assess how annotation quality influences model performance, we fine‑tuned the three best-performing pre-trained models (namely BOBs--LISA$_{+}$--\texttt{FT‑Real}, BOBs--LISA--\texttt{FT‑Real}, and BOBs--LISA--\texttt{Synth}) on three different subsets of the LISA training data: only \textit{good} annotations (based on our manual review) or the full set of \textit{all} annotations. \textit{Bad} annotations systematically led to poorer scores (\texttt{NormAvg} around 0.43--0.49) and were not reported. The results are summarized in Table~\ref{tab:goodbadall} and illustrated in Figure~\ref{fig:goodbadall}. Training on the subset with only good annotations consistently improved performance across both tasks and for nearly all model configurations. For example, the best single model for Task~2a (BOBs–LISA$_{+}$–\texttt{FT‑Real}) improved its \texttt{NormAvg} from 1.24 after pre‑training to 0.118 after fine‑tuning on only good images. A similar trend is observed in Task~2b, where the BOBs+LISA+\texttt{Synth} model improved from a \texttt{NormAvg} of 1.088 to 0.129 after fine‑tuning on good annotations. Generally, models fine‑tuned on only good annotations produced segmentations much more closely aligned with anatomical image features, as illustrated in Figure~\ref{fig:goodbadall}.

\subsection{Model ensembling}

\begin{figure}[t]
    \centering
    \includegraphics[width=1\linewidth]{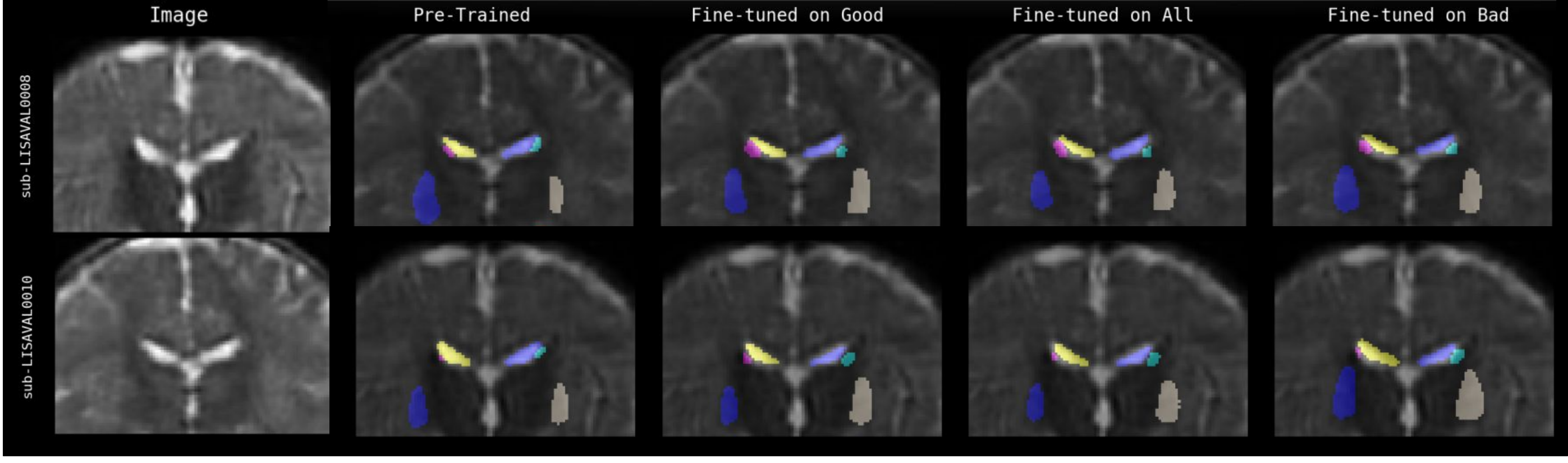}
    \caption{Effect of annotation quality on segmentation results on the BOBs--LISA model. Fine-tuning with only good annotations leads to clearly improved alignment with image features, whereas fine-tuning with bad or mixed annotations introduces misalignment due to registration errors, which are absent on the segmentations of the pre-trained model.}
    \label{fig:goodbadall}
    \captionof{table}{Recipes used for model ensembling. \textbf{Type} refers to the quality type of annotation used for fine-tuning (\texttt{all} samples or only the ones rated \texttt{good}), and \textbf{GT} refers to the labeling scheme used for the LISA data.}
    \label{tab:recipe}
    \resizebox{\linewidth}{!}{
    \begin{tabular}{p{1.2cm}ccccccccccc}
    \toprule
            \multirow{2}{*}{\parbox{1.2cm}{\textbf{Merged model}}}  & \multicolumn{3}{c}{\textbf{Model 1}} && \multicolumn{3}{c}{\textbf{Model 2}} & &\multicolumn{3}{c}{\textbf{Model 3}} \\
            \cmidrule(rl){2-4} \cmidrule(rl){6-8} \cmidrule(rl){10-12}
             & Pre-training & Type & GT && Pre-training & Type & GT && Pre-training & Type & GT \\
        \midrule
         \textbf{M1} &  BOBs+LISA+\texttt{Synth} & \texttt{all} & \texttt{GT}$_{\text{HF}}$ && BOBs+LISA+\texttt{FT-Real} & \texttt{good} & \texttt{GT}$_{\text{HF}}$ &&BOBs+LISA+\texttt{Synth} & \texttt{good} & \texttt{GT}$_{\text{HF}}$ \\
         \textbf{M2} &  BOBs+LISA+\texttt{Synth} & \texttt{all} & \texttt{GT}$_{\text{HF}}$ && BOBs+LISA+\texttt{FT-Real} & \texttt{good} & \texttt{GT}$_{\text{HF}}$ &&BOBs+LISA+\texttt{Synth} & \texttt{good} & \texttt{GT}$_{\text{HF}}$ \\
         \textbf{M3}& dHCP+LISA$_+$+\texttt{Synth} & \texttt{all} & \texttt{GT}$_{\text{LF}}$ && dHCP+\texttt{Synth}+LISA$_+$ & \texttt{good} & \texttt{GT}$_{\text{LF}}$  && BOBs+LISA$_+$+\texttt{FT-Real} & \texttt{good} & \texttt{GT}$_{\text{HF}}$ \\
         \textbf{M4}& \multicolumn{3}{c}{--- Model \textbf{M1} --- }&& \multicolumn{3}{c}{--- Model \textbf{M2} --- } &&dHCP+LISA$_+$+\texttt{Synth} & \texttt{all} & \texttt{GT}$_{\text{LF}}$ \\
         \bottomrule
         
    \end{tabular}}
\vspace{-.3cm}
\end{figure}

In our final set of experiments, we explored various ensembling strategies. For these experiments, we selected the three best models trained on the BOBs dataset with LISA annotations for each task. We also leveraged the two sets of available annotations for the hippocampi and basal ganglia, namely the ones done on ULF (\texttt{GT}$_{\text{ULF}}$) and the ones done on HF and propagated to ULF (\texttt{GT}$_{\text{HF}}$), used for the evaluation. The recipes of models used are presented in Table~\ref{tab:recipe}. \textbf{M1} and \textbf{M2} contain the top-three individual models for tasks 2a and 2b. \textbf{M3} and \textbf{M4} were built less systematically, through trial of various combinations.

\begin{table}[h]
\vspace{-.5cm}
\centering
\setlength{\tabcolsep}{4pt}
\caption{Comparison of ensembling strategies defined in Table~\ref{tab:recipe} and best individual models (resp. BOBs+LISA$_+$+\texttt{FT-Real} and BOBs+LISA+\texttt{Synth}). }
\label{tab:ensemble_best}
\resizebox{.9\linewidth}{!}{
\begin{tabular}{c l c c c c c c}
\toprule
\rotatebox{0}{Task}
  & \rotatebox{0}{Strategy}
    & \texttt{NormAvg} $\downarrow$ & DSC $\uparrow$ & HD $\downarrow$ & HD95 $\downarrow$ & ASSD $\downarrow$ & RVE $\downarrow$ \\
\midrule
\multirow{6}{*}{\rotatebox[origin=c]{90}{\textsc{2a--Hippo.}}}
  & Best indiv. 2a   &   {0.118} & 0.69{\tiny ±0.18} & 5.73{\tiny ±7.79} & 2.02{\tiny ±1.42} & 0.80{\tiny ±0.90} & 0.12{\tiny ±0.08} \\
  & Best indiv. 2b     & {0.183}& 0.70{\tiny ±0.19}& 6.06{\tiny ±7.74}& 2.23{\tiny ±1.56}& 0.81{\tiny ±0.95}& 0.18{\tiny ±0.07}\\
  \cmidrule{2-8}
  & \textbf{M1}      & 0.141 & 0.70{\tiny ±0.19} & 5.79{\tiny ±7.91} & 2.11{\tiny ±1.41} & 0.77{\tiny ±0.88} & 0.16{\tiny ±0.08} \\
  & \textbf{M2}      & 0.161 & 0.69{\tiny ±0.19} & 6.08{\tiny ±8.06} & 2.15{\tiny ±1.48} & 0.80{\tiny ±0.94} & 0.14{\tiny ±0.08} \\
   & \textbf{M3} & \textbf{0.075} & 0.71{\tiny ±0.19} & 5.65{\tiny ±7.98} & 1.95{\tiny ±1.49} & 0.76{\tiny ±0.96} & 0.13{\tiny ±0.07} \\
  & \textbf{M4}  & 0.125 & 0.70{\tiny ±0.19} & 5.90{\tiny ±7.94} & 2.07{\tiny ±1.48} & 0.77{\tiny ±0.89} & 0.14{\tiny ±0.07} \\
\midrule
\multirow{6}{*}{\rotatebox[origin=c]{90}{\textsc{2b--BaGa}}}
   & Best indiv. 2a     & 0.196& 0.85{\tiny ±0.07}& 3.28{\tiny ±1.19}& 1.89{\tiny ±1.05}& 0.52{\tiny ±0.29} & 0.08{\tiny ±0.03}\\
  & Best indiv. 2b     & {0.129}& 0.86{\tiny ±0.04}& 3.04{\tiny ±0.69}& 1.64{\tiny ±0.69}& 0.48{\tiny ±0.21}& 0.09{\tiny ±0.03}\\
  \cmidrule{2-8}
  & \textbf{M1}        & 0.076 & 0.86{\tiny ±0.05} & 2.92{\tiny ±0.84} & 1.59{\tiny ±0.79} & 0.45{\tiny ±0.24} & 0.07{\tiny ±0.02} \\

  & \textbf{M2}        & 0.065 & 0.86{\tiny ±0.05} & 3.02{\tiny ±0.86} & 1.61{\tiny ±0.83} & 0.44{\tiny ±0.23} & 0.06{\tiny ±0.03} \\

   & \textbf{M3}   & 0.093 & 0.86{\tiny ±0.06} & 2.96{\tiny ±1.10} & 1.69{\tiny ±0.91} & 0.46{\tiny ±0.26} & 0.07{\tiny ±0.03} \\
  & \textbf{M4}   & \textbf{0.032} & 0.87{\tiny ±0.05} & 3.01{\tiny ±0.82} & 1.56{\tiny ±0.77} & 0.44{\tiny ±0.22} & 0.06{\tiny ±0.03} \\
\bottomrule
\end{tabular}}
\vspace{-.3cm}
\end{table}

Results are provided in Table~\ref{tab:ensemble_best}.  Ensembling improves performance compared to individual models. M3 and M4 achieve the strongest performances on Task 2a and 3b, respectively, outperforming the more homogeneous approaches of M1 and M2. M3 and M4, respectively, achieved a \texttt{NormAvg} of 0.0754 for Task~2a and of 0.0319 for Task~2b.  We also observed that models trained using \texttt{GT}$_{\text{LF}}$ annotations consistently outperformed those trained on \texttt{GT}$_{\text{HF}}$ annotations, even when evaluated on them. We believe that this could warrant further investigation for a future edition of the challenge.

\vspace{-.3cm}
\section{Discussion}
This study shows that domain randomization, combined with careful data curation and model fusion, can effectively bridge the gap between HF and ULF MRI for neonatal brain segmentation and, in some cases, work better for annotation transfer from HF to ULF than simple subject co‑registration approaches.

Domain randomization significantly improves generalization: BOBs-pretrained models allowed to achieve strong transfer performance when fine-tuned on LISA data. Both LISA and LISA$_{+}$ annotations were useful, with LISA$_{+}$ performing best on Task~2a and LISA-annotations on Task~2b. 

An important advantage of domain randomization is that, since images are generated directly from segmentation maps, the resulting image–annotation pairs are perfectly aligned, with image contrasts matching annotation borders -- something difficult to achieve with manual annotations. Annotation quality proved critical: models trained on poorly aligned annotations overfit to registration errors, while using only high‑quality annotations improved performance and alignment with image features. HF annotations transferred to LF images via registration were unreliable in around 30\% of the data, and excluding them from training helped improve performance. 

Finally, model ensembling further boosted performance. As commonly observed in medical imaging challenges~\cite{eisenmann2023winner}, a highly diverse, heuristically built ensemble of models trained throughout our experiments was found to achieve the best performance, although this approach might not generalize as well to new tasks as more principled ensembles. 

In this work, we illustrated the strength of domain‑randomization-based approaches for HF-to-ULF knowledge transfer. We believe that these methods could be instrumental in designing robust and generalizable models for ULF pediatric segmentation tasks.

\vspace{-.3cm}

\begin{credits}
\subsubsection{\ackname} This research was funded by the Swiss National Science Foundation (215641), ERA-NET Neuron MULTI-FACT project (SNSF 31NE30\_203977); we acknowledge the Leenaards and Jeantet Foundations as well as CIBM Center for Biomedical Imaging, a Swiss research center of excellence founded and supported by CHUV, UNIL, EPFL, UNIGE and HUG. This research was also supported by grants from NVIDIA and utilized NVIDIA RTX6000 ADA GPUs. The Developing Human Connectome Project (dHCP) was funded by the European Research Council (ERC) under the European Union’s Seventh Framework Programme (FP7/2007–2013), Grant Agreement No. 319456.

\subsubsection{\discintname}
The authors have no competing interests to disclose.
\end{credits}


\printbibliography
\appendix
\makeatletter
\setcounter{table}{0}
\setcounter{figure}{0}
\renewcommand 
\thesection{S\@arabic\c@section}
\renewcommand\thetable{S\@arabic\c@table}
\renewcommand \thefigure{S\@arabic\c@figure}
\makeatother
\newpage
{\Large \noindent \textbf{Supplementary material}}
\section{Manual label quality evaluation}\label{app:qc}
\begin{figure}
    \centering
    \includegraphics[width=\linewidth]{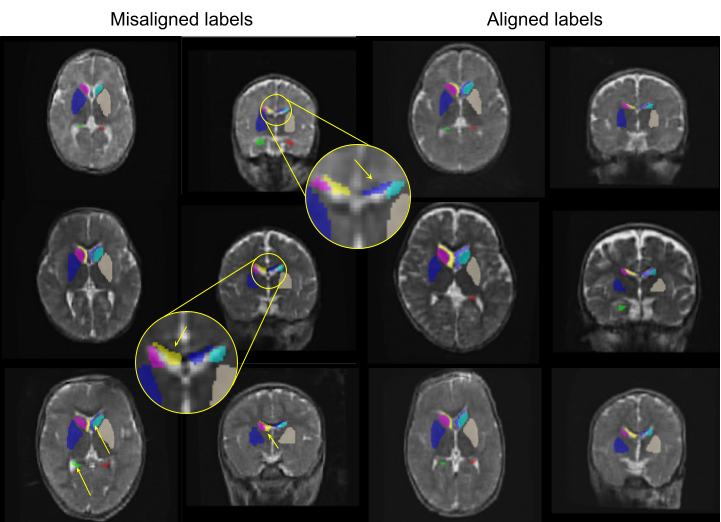}
    \caption{\textbf{A more detailed outlook at our manual quality annotations.} The left columns feature 3 subjects with misaligned labels, and the right column features three different subjects with well-aligned labels.}
    \label{fig:annotations}
\end{figure}

We provide here a more detailed outlook at our manual quality assessment. We carried out a visual qualitative assessment of label alignment across all available data, focusing around the coronal view of the images, using ITK-Snap. As illustrated in Figure~\ref{fig:annotations}, the coronal view revealed several subjects with the caudate and ventricles being shifted. This is particularly visible in the zoomed-in subjects, where the left hemisphere ventricle (first row)  and the right hemisphere ventricle (second row) are shifted and do not cover the ventricle (bright tissue on the image). These data were subsequently excluded from training, resulting in the discarding of 23 our of 79 data points. The complete manual rating is available at \url{https://github.com/Medical-Image-Analysis-Laboratory/lisasegm/blob/main/LISA_QC.csv}.

\end{document}